\documentclass[journal]{IEEEtran}
\IEEEoverridecommandlockouts
\usepackage{lineno}
\usepackage{hyperref}
\usepackage{cite}
\usepackage{amsmath,amssymb,amsfonts,cases} 
\usepackage{amsmath}
\usepackage{amsthm}
\DeclareMathOperator*{\argmax}{argmax}

\usepackage{graphicx}
\usepackage{textcomp}
\usepackage{xcolor}
\usepackage{graphicx}
\usepackage{float}
\usepackage{subfigure}
\usepackage{amsmath,mleftright}
\usepackage{amsfonts,amssymb}
\usepackage{mathrsfs}
\usepackage{mathtools}
\usepackage{algorithm}
\usepackage{algorithmic}
\usepackage{bm}
\usepackage{multirow}
\usepackage{array}
\usepackage{amssymb}
\usepackage{amsmath}
\usepackage{cite}
\usepackage{url}
\usepackage{xcolor}
\usepackage{cite,graphicx,amsmath,amssymb}
\usepackage{subfigure}
\usepackage{fancyhdr}
\usepackage{mdwmath}
\usepackage{mdwtab}
\usepackage{caption}
\usepackage{amsthm}
\usepackage{setspace}
\usepackage{bm}
\usepackage{mathtools}
\usepackage{dsfont}
\usepackage{bbm}
\usepackage{framed}

\newtheorem{theorem}{Theorem}

\newtheorem{lemma}{Lemma}

\newtheorem{corollary}{Corollary}

\makeatletter
\newcommand{\biggg}{\bBigg@{3}}
\newcommand{\Biggg}{\bBigg@{3.5}}
\makeatother
\makeatletter
\allowdisplaybreaks[4]
\renewcommand{\maketag@@@}[1]{\hbox{\m@th\normalsize\normalfont#1}}%
\makeatother
\def\BibTeX{{\rm B\kern-.05em{\sc i\kern-.025em b}\kern-.08em
    T\kern-.1667em\lower.7ex\hbox{E}\kern-.125emX}}
    \expandafter\def\expandafter\normalsize\expandafter{%
    \normalsize%
    \setlength\abovedisplayskip{4pt}%
    \setlength\belowdisplayskip{4pt}%
    \setlength\abovedisplayshortskip{2pt}%
    \setlength\belowdisplayshortskip{2pt}%
}
\begin{document}
\title{Joint Segment Activation and Antenna Placement for Uplink SWAN Systems}
\author{Songnan~Gu, Zhenqiao~Cheng, Hao~Jiang, Chongjun~Ouyang,\\ Yuanwei~Liu, \IEEEmembership{Fellow, IEEE}, and Arumugam~Nallanathan, \IEEEmembership{Fellow, IEEE}\vspace{-10pt}
\thanks{S. Gu, H. Jiang, C. Ouyang, and A. Nallanathan are with the School of Electronic Engineering and Computer Science, Queen Mary University of London, London E1 4NS, U.K. (e-mail: s.gu@se23.qmul.ac.uk; \{hao.jiang, c.ouyang, a.nallanathan\}@qmul.ac.uk).}
\thanks{Z. Cheng is with the 6G Research Centre, China Telecom Beijing Research Institute, Beijing 102209, China (e-mail: zhenqiao.cheng@engineer.com).}
\thanks{Y. Liu is with the Department of Electrical and Electronic Engineering, The University of Hong Kong, Hong Kong (e-mail: yuanwei@hku.hk).}
}
\maketitle
\begin{abstract}
This article analyzes the achievable sum-rate of multiuser uplink segmented waveguide-enabled pinching-antenna systems (SWANs). To unveil system-design insights, an upper bound on the achievable sum-rate is derived, based on which the existence of an optimal segment activation level is theoretically established. Motivated by this result, hybrid segment selection and aggregation (HSS/A) schemes are proposed to jointly optimize segment activation and pinching-antenna (PA) placement. Correspondingly, low-complexity greedy algorithms are developed for the considered optimization problem. Numerical results validate the theoretical analysis and demonstrate that the proposed HSS/A schemes outperform conventional full-segment aggregation.
\end{abstract}
\section{Introduction}
The pinching-antenna system (PASS) has recently attracted significant attention due to their capability to mitigate large-scale path loss and blockage effects through waveguide-assisted signal transmission \cite{suzuki2022pinching,Ding2025FlexibleAntenna}. This feature makes PASS a promising architecture for high-frequency wireless communications, especially for the emerging upper-mid-band and millimeter-wave systems. In PASS, electromagnetic signals are guided through low-loss dielectric waveguides and are transmitted or captured through small dielectric pinching antennas (PAs) attached to the waveguide. Since the PAs can be flexibly deployed along the waveguide, PASS enables meter-scale spatial reconfiguration and can effectively shorten the free-space propagation distance \cite{suzuki2022pinching,Ding2025FlexibleAntenna}.

Despite these advantages, uplink PASS still faces a fundamental challenge caused by inter-antenna radiation (IAR). In conventional uplink PASS, multiple PAs are simultaneously deployed on the same dielectric waveguide to receive uplink signals from different users \cite{tegos2024minimum,Lyu2025PASSAirComp,Ouyang2026SWAN}. Since each PA continuously leaks electromagnetic energy into the waveguide, the received signals may be re-radiated by other PAs before reaching the feed point. This effect introduces strong coupling among different PAs and substantially complicates the uplink signal model and receiver design \cite{Ouyang2026SWAN}.

\begin{figure}[!t]
\centering
    \subfigure[Segmented waveguide.]
    {
        \includegraphics[width=0.35\textwidth]{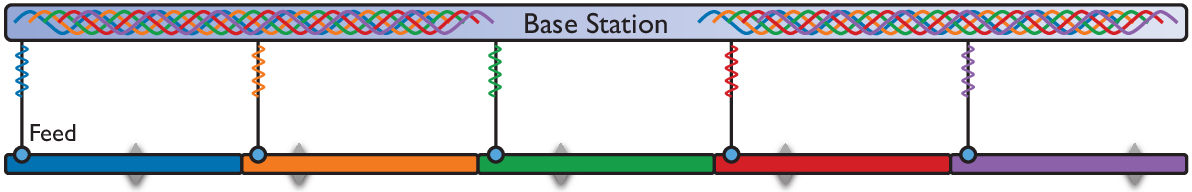}
	   \label{Figure: PAS_System_Model2}
    }
    \subfigure[System setup.]
    {
        \includegraphics[width=0.35\textwidth]{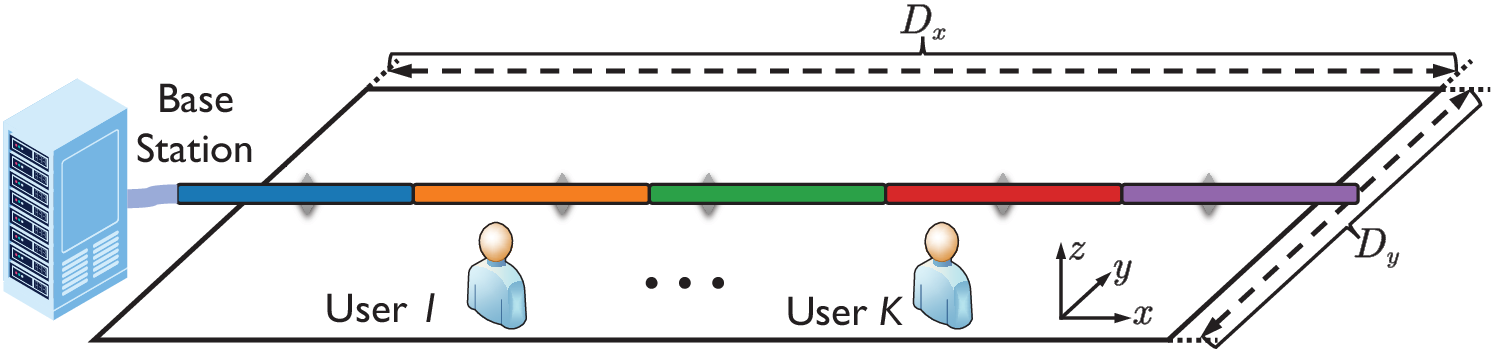}
	   \label{Figure: PAS_System_Model1}
    }
\caption{Illustration of the SWAN-based uplink channel.}
\label{Figure: PAS_System_Model}
\vspace{-15pt}
\end{figure}

To overcome this issue, the segmented waveguide-enabled pinching-antenna system (SWAN) was recently proposed in \cite{Ouyang2026SWAN}. In SWAN, the long dielectric waveguide is divided into multiple independent short segments, where each segment is connected to the base station (BS) through an individual feed point; see {\figurename} {\ref{Figure: PAS_System_Model}}. Therefore, at most one PA is activated in each segment, which completely eliminates IAR and provides a more tractable uplink transmission architecture.

Although SWAN-based transmission has recently attracted increasing research interest, the corresponding studies are still in their infancy. In our previous works \cite{Ouyang2026SWAN,gu2025sum}, the uplink sum-rate maximization problem for SWAN was investigated. Numerical results revealed an interesting non-monotonic phenomenon: \emph{activating more segments does not always improve the achievable sum-rate}, although additional segments are intuitively expected to provide stronger array gain. However, this counter-intuitive phenomenon has not yet been theoretically characterized, and no transmission strategy has been developed to explicitly exploit it in system design.

To fill this gap, this article investigates uplink sum-rate maximization for multiuser SWAN systems. We first derive an upper bound on the achievable sum-rate and theoretically prove the existence of an optimal segment activation level, which results from the tradeoff between coherent signal aggregation and accumulated receiver noise. Motivated by this observation, we propose a hybrid segment selection and aggregation (HSS/A) architecture, where only a subset of segments is activated for uplink signal aggregation. We further develop low-complexity greedy algorithms to jointly optimize segment activation and PA placement. Numerical results validate the theoretical analysis and demonstrate that the proposed HSS/A schemes outperform conventional full-segment aggregation.

\section{System Model}
We consider a multiuser uplink SWAN system where a BS serves $K$ users distributed within a rectangular area of dimensions $D_x$ and $D_y$ along the $x$- and $y$-axes, respectively. Let $\mathcal K\triangleq\{1,\ldots,K\}$ denote the user set. The coordinate of user $k$ is denoted by $\mathbf u_k=[u_k^x,u_k^y,0]^{\mathsf{T}}$. The SWAN architecture consists of $M$ dielectric waveguide segments, each with length $L$, such that $ML\geq D_x$, which is illustrated in {\figurename} {\ref{Figure: PAS_System_Model}}. The segments are deployed along the $x$-axis at height $d$. For segment $m\in\mathcal M\triangleq\{1,\ldots,M\}$, the feed point is denoted by $\boldsymbol{\psi}_0^m=[\psi_0^m,0,d]^{\mathsf{T}}$, where $\psi_0^1<\psi_0^2<\cdots<\psi_0^M$. Each feed point is located at the beginning of its corresponding segment.

To eliminate IAR, at most one PA is activated in each segment. The coordinate of the active PA in segment $m$ is denoted by
$\boldsymbol{\psi}_m=[\psi_m,0,d]^{\mathsf{T}}$.
The PA locations satisfy
\begin{align}
\psi_0^m
\le
\psi_m
\le
\psi_0^m+L,
|\psi_m-\psi_{m'}|
\ge
\Delta,
\forall m\neq m',
\end{align}
where $\Delta>0$ denotes the minimum inter-antenna spacing.
\subsection{Channel Model}
PASS is envisioned for operation in high-frequency bands \cite{suzuki2022pinching}, where line-of-sight (LoS) propagation dominates \cite{ouyang2024primer}. Therefore, a free-space LoS channel model is adopted. The channel coefficient between user $k$ and the PA in segment $m$ can be characterized as follows \cite{ouyang2024primer}:
\begin{align}
    h_{\rm{o}}({\mathbf{u}}_k,{\bm\psi}_{m})\triangleq
\frac{\eta^{\frac{1}{2}}{\rm{e}}^{-{\rm{j}}k_0\lVert{\mathbf{u}}_k-{\bm\psi}_{m}\rVert}}{\lVert{\mathbf{u}}_k-{\bm\psi}_{m}\rVert},
\end{align}
where $\eta=\frac{c^2}{16\pi^2f_c^2}$, $c$ denotes the speed of light, $f_c$ denotes the carrier frequency, $\lambda$ denotes the free-space wavelength, and $k_0=\frac{2\pi}{\lambda}$ denotes the wavenumber. The in-waveguide propagation coefficient between the feed point and the PA in segment $m$ is modeled as follows \cite{pozar2021microwave}:
\begin{align}
   h_{\rm{i}}({\bm\psi}_{m},{\bm\psi}_{0}^{m})\triangleq{10^{-\frac{\kappa}{20}\lVert{\bm\psi}_{m}-{\bm\psi}_{0}^{m}\rVert}}
{\rm{e}}^{-{\rm{j}}\frac{2\pi\lVert{\bm\psi}_{m}-{\bm\psi}_{0}^{m}\rVert}{\lambda_{\rm{g}}}},
\end{align}
where
$\lambda_g=\frac{\lambda}{n_{\rm eff}}$, $n_{\rm eff}$ denotes the effective refractive index of the dielectric waveguide, and $\kappa$ denotes the in-waveguide attenuation factor. Prior studies showed that the in-waveguide attenuation has negligible impact on practical SWAN systems \cite{Ouyang2026SWAN}. Therefore, we set $\kappa=0$ in this work.

\begin{figure}[!t]
\centering
    \subfigure[Type-I SA.]
    {
        \includegraphics[width=0.2\textwidth]{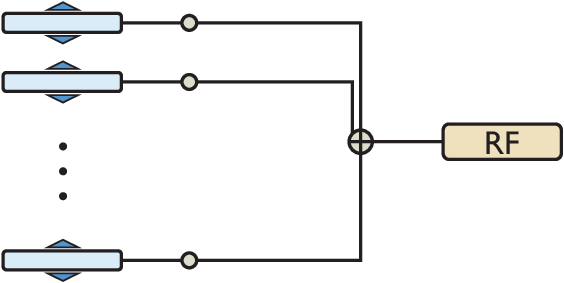}
	   \label{Figure: PAN_Protocol2}
    }
    \subfigure[Type-II SA.]
    {
        \includegraphics[width=0.2\textwidth]{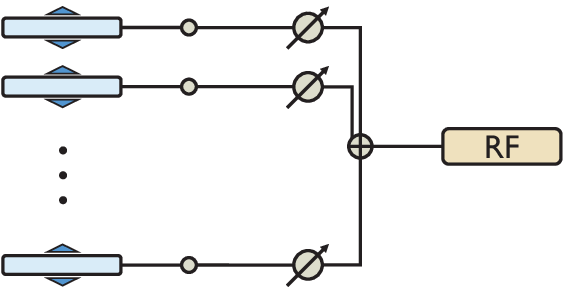}
	   \label{Figure: PAN_Protocol5}
    }
\caption{Illustration of the SA-based SWAN architectures.}
\label{Figure: PAN_Protocol}
\vspace{-15pt}
\end{figure}

\subsection{Segment Aggregation Architectures}
The performance of SWAN depends on how the segment feed points are connected to the RF chain. To exploit the spatial array gain provided by multiple segments, This work considers two representative SA architectures.
\subsubsection{Type-I SA}
In Type-I SA, shown in Fig.~\ref{Figure: PAN_Protocol2}, all segment feed points are directly connected to a single RF chain through an analog combiner. The effective uplink channel of user $k$ can be written as follows:
\begin{align}
h_k&=
\frac{1}{\sqrt{M}}
\sum_{m=1}^{M}
h_{\rm i}({\bm\psi}_m,{\bm\psi}_0^m)
h_{\rm o}({\mathbf u}_k,{\bm\psi}_m)\\
&=\frac{\eta^{\frac{1}{2}}}{\sqrt{M}}
\sum_{m=1}^{M}\frac{{\rm{e}}^{-{\rm{j}}k_0(\sqrt{(u_k^x-{\psi}_{m})^2+d_k}+n_{\rm{eff}}({\psi}_{m}-{\psi}_{0}^{m})) }}{\sqrt{(u_k^x-{\psi}_{m})^2+d_k}},\label{Channel_Type_I_SA}
\end{align}
where $d_k\triangleq d^2+(u_k^y)^2$. The factor $\frac1{\sqrt M}$ accounts for the cumulative receiver noise introduced by analog signal combining.
\subsubsection{Type-II SA}
To further improve coherent signal combining, we also consider Type-II SA, shown in Fig.~\ref{Figure: PAN_Protocol5}, where each segment is additionally equipped with one analog phase shifter before RF combining.

The effective uplink channel of user $k$ becomes
\begin{align}
h_k
=
\frac{1}{\sqrt{M}}
\sum_{m=1}^{M}
{\rm e}^{{\rm j}\theta_m}
h_{\rm i}({\bm\psi}_m,{\bm\psi}_0^m)
h_{\rm o}({\mathbf u}_k,{\bm\psi}_m),
\end{align}
where
$\theta_m\in[0,2\pi)$
denotes the phase shift applied to segment $m$. Substituting the channel expressions yields
\begin{align}
h_k
=
\sum_{m=1}^{M}\frac{\eta^{\frac{1}{2}}{\rm{e}}^{{\rm j}\theta_m-{\rm{j}}k_0(\sqrt{(u_k^x-{\psi}_{m})^2+d_k}+n_{\rm{eff}}({\psi}_{m}-{\psi}_{0}^{m})) }}{\sqrt{M}\sqrt{(u_k^x-{\psi}_{m})^2+d_k}}.\label{Channel_Type_II_SA}
\end{align}
Type-I SA can be regarded as a special case of Type-II SA with $\theta_m=0$ for all $m$.
\subsection{Signal Model}
The received signal at the BS is given by
\begin{align}
y
=
\sum_{k=1}^K
\sqrt{P_k}
h_k
x_k
+n,
\end{align}
where
$x_k$ denotes the normalized data symbol of user $k$ satisfying
$\mathbbmss E[x_k]=0$
and
$\mathbbmss E[|x_k|^2]=1$,
$P_k$ denotes the transmit power of user $k$,
and
$n\sim\mathcal{CN}(0,\sigma^2)$
denotes additive white Gaussian noise with noise power $\sigma^2$. The achievable sum-rate is given by
\begin{equation}\label{Rate_Uplink}
    {\mathcal{R}} = \log_2 \left( 1 + \sum\nolimits_{k=1}^{K} {P_k}\lvert h_k\rvert^2/{\sigma^2} \right).
\end{equation}
In the sequel, we analyze the achievable sum-rate and reveal the existence of an optimal number of activated segments.

\section{Sum-Rate Analysis}\label{Section: Sum-Rate Analysis}
In this section, we analyze the achievable sum-rate of the considered SA architectures. Direct analysis of \eqref{Rate_Uplink} is difficult because the PA locations, phase shifts, and user-dependent path losses are strongly coupled. To obtain useful insights, we derive an upper bound on the per-user channel gain under ideal coherent combining. The derived bound reveals that full-segment activation is generally suboptimal and that an optimal number of activated segments exists.

For notational simplicity, define the cascaded channel between user $k$ and the PA in segment $m$ as follows:
\begin{align}
g_{k,m}(\psi_m)
\triangleq
h_{\rm i}(\boldsymbol{\psi}_m,\boldsymbol{\psi}_0^m)
h_{\rm o}(\mathbf u_k,\boldsymbol{\psi}_m).
\end{align}
By the triangle inequality, the effective channel gain satisfies
\begin{align}
|h_k|^2
\leq
\frac{1}{M}
\left(
\sum_{m=1}^{M}
|g_{k,m}(\psi_m)|
\right)^2.
\label{Channel_Bound_Triangle}
\end{align}
Equality in \eqref{Channel_Bound_Triangle} is achieved only when all received components are perfectly phase-aligned. Therefore, the bound corresponds to ideal coherent combining. It applies to Type-II SA with optimal phase shifts and also upper bounds Type-I SA. We next upper bound the right-hand side of \eqref{Channel_Bound_Triangle}. For user $k$, the strongest contribution is achieved when one PA is placed at the projection of the user onto the waveguide. The corresponding channel amplitude satisfies
\begin{align}
|g_{k,m_k}|
\leq
\frac{\eta^{\frac12}}{\sqrt{d_k}},
\end{align}
where $d_k \triangleq d^2+(u_k^y)^2$, and $m_k$ denotes the index of the segment containing the projection of user $k$.

The remaining activated segments are distributed on the left and right sides of the user projection. Let $M_k^{-}\triangleq m_k-1$ and $M_k^{+}\triangleq M-m_k$ denote the numbers of activated segments on the left and right sides, respectively, such that 
\begin{align}
M=M_k^{-}+M_k^{+}+1. 
\end{align}
Moreover, let $\delta_k^{-}\in[0,L]$ and $\delta_k^{+}\in[0,L]$ denote the horizontal distances from the user projection to the nearest left and right segment edges, respectively. By neglecting the minimum-spacing constraint and placing each PA at the closest point to the user projection within its segment, we obtain
\begin{align}
\sum_{m=1}^{M}
|g_{k,m}(\psi_m)|
\leq
\eta^{\frac12}
\left[
\frac{1}{\sqrt{d_k}}
+
f_k(\delta_k^{-},M_k^{-})
+
f_k(\delta_k^{+},M_k^{+})
\right],
\label{Amplitude_Bound_Sum}
\end{align}
where $
f_k(\delta,N)
\triangleq
\sum_{n=1}^{N}
\frac{1}{\sqrt{(\delta+(n-1)L)^2+d_k}}$. The bound in \eqref{Amplitude_Bound_Sum} is optimistic because the mutual coupling and minimum-spacing constraints are ignored. Therefore, it is suitable for deriving a performance upper bound.

For large $M_k^{-}$ and $M_k^{+}$, the two summations in
\eqref{Amplitude_Bound_Sum} can be accurately approximated
by midpoint Riemann integrals. Recall that $f_k(\delta,N)
\triangleq
\sum_{n=1}^{N}
\frac{1}{\sqrt{(\delta+(n-1)L)^2+d_k}}$, where $\delta\in[0,L]$ denotes the horizontal distance between
the user projection and the nearest segment edge. The $n$th term in the above summation corresponds to the
midpoint sample over the interval
$
[\delta+(n-\frac32)L,\,
\delta+(n-\frac12)L]
$.
Therefore, $f_k(\delta,N)$ can be approximated as follows:
\begin{align}
f_k(\delta,N)
&\approx
\frac{1}{L}
\int_{\max\{\delta-\frac{L}{2},0\}}^{\delta+(N-\frac12)L}
\frac{1}{\sqrt{x^2+d_k}}
\,dx \\
&=
\frac{1}{L}
\Bigg[
\sinh^{-1}
\left(
\frac{\delta+(N-\frac12)L}{\sqrt{d_k}}
\right)
\nonumber\\
&
-
\sinh^{-1}
\left(
\frac{\max\{\delta-\frac{L}{2},0\}}{\sqrt{d_k}}
\right)
\Bigg]\triangleq \Phi_k(N,\delta).
\end{align}

Therefore, the per-user channel gain is upper bounded by
\begin{align}
|h_k|^2
&\leq
\frac{\eta}{M}
\left[
\frac{1}{\sqrt{d_k}}
+
\frac{\Phi_k(M_k^{-},\delta_k^{-})}{L}
+
\frac{\Phi_k(M_k^{+},\delta_k^{+})}{L}
\right]^2
\label{User_Channel_Gain_Upper_Bound}\\
&\triangleq
\overline G_k(M,M_k^{-},M_k^{+}).
\end{align}
Using \eqref{User_Channel_Gain_Upper_Bound}, the uplink sum-rate is upper bounded by
\begin{align}
\mathcal R
\leq
\log_2
\left(
1+
\frac1{\sigma^2}
\sum_{k=1}^{K}
P_k
\overline G_k(M,M_k^{-},M_k^{+})
\right).
\label{Sum_Rate_Upper_Bound}
\end{align}

We next examine the asymptotic behavior of the bound. Three cases are considered.
\subsubsection*{Case I}
$M_k^{-}\rightarrow\infty$ and $M_k^{+}\rightarrow\infty$. In this case, $M\rightarrow\infty$. Since $\sinh^{-1}(x)$ grows only logarithmically with $x$, the bracketed term in \eqref{User_Channel_Gain_Upper_Bound} scales no faster than $\mathcal O(\log M)$. Consequently, the squared term scales as $\mathcal O((\log M)^2)$, whereas the denominator grows linearly with $M$. Therefore,
\begin{align}
\lim_{M\rightarrow\infty}
\overline G_k(M,M_k^{-},M_k^{+})
=0.
\end{align}
\subsubsection*{Case II}
$M_k^{-}$ is finite while $M_k^{+}\rightarrow\infty$. In this case, only one side contributes an unbounded logarithmic term. Nevertheless, the denominator still grows linearly with $M$. It follows that
\begin{align}
\lim_{M\rightarrow\infty}
\overline G_k(M,M_k^{-},M_k^{+})
=0.
\end{align}
\subsubsection*{Case III}
$M_k^{-}\rightarrow\infty$ while $M_k^{+}$ is finite. This case is symmetric to Case II, and we similarly obtain
\begin{align}
\lim_{M\rightarrow\infty}
\overline G_k(M,M_k^{-},M_k^{+})
=0.
\end{align}

Therefore, for all three asymptotic cases,
\begin{align}
\lim_{M\rightarrow\infty}
\overline G_k(M,M_k^{-},M_k^{+})
=0,
\quad
\forall k\in\mathcal K.
\end{align}
Substituting the above result into \eqref{Sum_Rate_Upper_Bound} yields
\begin{align}
\lim_{M\rightarrow\infty}
\mathcal R
=0.
\end{align}

Since the achievable sum-rate is strictly positive for any finite feasible segment configuration with nonzero channel gains, there exists at least one finite number of activated segments that maximizes the achievable sum-rate. This result shows that full-segment activation is generally suboptimal. The reason is that the additional array gain contributed by distant segments increases only logarithmically, whereas the accumulated receiver noise increases linearly with the number of activated segments. This tradeoff motivates the joint segment activation and PA placement methods developed in the next section.

\begin{figure}[!t]
\centering
    \subfigure[Type-I HSS/A.]
    {
        \includegraphics[width=0.2\textwidth]{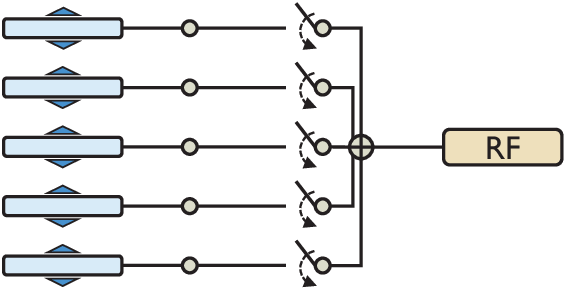}
	   \label{Figure: PAN_7}
    }
    \subfigure[Type-II HSS/A.]
    {
        \includegraphics[width=0.2\textwidth]{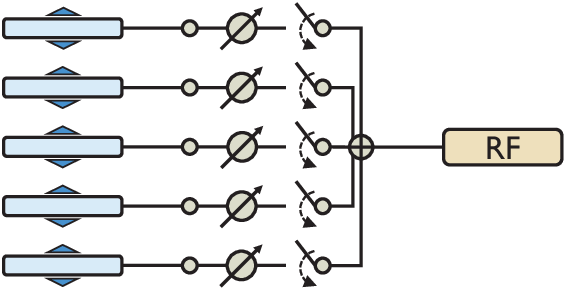}
	   \label{Figure: PAN_6}
    }
\caption{Illustration of the HSS/A architectures.}
\label{Figure: PAN}
\vspace{-15pt}
\end{figure}

\section{Joint Segment Activation and Antenna Placement}
The analysis in the previous section shows that full-segment activation is generally suboptimal. This motivates a HSS/A architecture, where only a subset of segments is activated and aggregated. As illustrated in Fig.~\ref{Figure: PAN}, two HSS/A architectures are considered.
\begin{itemize}
\item \emph{Type-I HSS/A}: each segment is connected to the combiner through a switch. Only segment activation and PA placement are optimized.

\item \emph{Type-II HSS/A}: each segment is additionally equipped with one analog phase shifter before signal combining. Segment activation, PA placement, and phase shifts are jointly optimized.
\end{itemize}
Let $\mathcal S\subseteq\mathcal M$ denote the activated segment set and let $S=|\mathcal S|$. For Type-I HSS/A, the effective channel of user $k$ is
\begin{align}
h_k(\mathcal S,\boldsymbol{\psi})
=
\frac{1}{\sqrt S}
\sum_{m\in\mathcal S}
g_{k,m}(\psi_m).
\end{align}
For Type-II HSS/A, the effective channel becomes
\begin{align}
h_k(\mathcal S,\boldsymbol{\psi},\boldsymbol{\theta})
=
\frac{1}{\sqrt S}
\sum_{m\in\mathcal S}
{\rm {e}}^{{\rm{j}}\theta_m}
g_{k,m}(\psi_m),
\end{align}
where $\theta_m\in[0,2\pi)$ is the phase shift applied to segment $m$.
\subsection{Type-I HSS/A}
For a given activated segment set $\mathcal S$, the feasible PA placement set is given by
\begin{align}
\mathcal X(\mathcal S)
=
\left\{
\boldsymbol{\psi}
\left|
\psi_m\in[\psi_0^m,\psi_0^m+L],
~\forall m\in\mathcal S
\right.
\right\}.
\end{align}
The Type-I HSS/A design problem is formulated as follows:
\begin{align}
\max_{\mathcal S\subseteq\mathcal M,\,
\boldsymbol{\psi}\in\mathcal X(\mathcal S)}
\mathcal R(\mathcal S,\boldsymbol{\psi}).
\end{align}
The above problem is difficult to solve optimally because it involves both discrete segment activation and continuous PA placement.

To obtain a low-complexity solution, we adopt a greedy activation strategy with full-level evaluation. At iteration $i$, the current activated segment set is denoted by $\mathcal S_{i-1}$. For each inactive segment $m\in\mathcal M\setminus\mathcal S_{i-1}$, segment $m$ is temporarily activated and its PA position is optimized while all previously selected PA positions remain fixed.

The feasible interval of $\psi_m$ is discretized as follows:
\begin{align}
\mathcal Q_m
=
\left\{
\left.
\psi_0^m+\frac{qL}{Q-1}
\right|
q=0,1,\ldots,Q-1
\right\}.
\end{align}
For a current segment set $\mathcal S$, the infeasible grid points caused by the spacing constraint are given by
\begin{align}
\hat{\mathcal Q}_m(\mathcal S)
=
\left\{
x\in\mathcal Q_m
\left|
|x-\psi_{m'}|<\Delta,
~\exists m'\in\mathcal S
\right.
\right\}.
\end{align}
For candidate segment $m$, the trial PA position is selected as follows:
\begin{align}
\psi_m^\star
=
\argmax_{\psi_m\in
\mathcal Q_m\setminus
\hat{\mathcal Q}_m(\mathcal S_{i-1})}
\mathcal R
\left(
\mathcal S_{i-1}\cup\{m\},
\boldsymbol{\psi}_{i-1}\cup\{\psi_m\}
\right).
\end{align}
The segment added at iteration $i$ is determined as follows:
\begin{align}
m_i^\star
=
\argmax_{m\in\mathcal M\setminus\mathcal S_{i-1}}
\mathcal R
\left(
\mathcal S_{i-1}\cup\{m\},
\boldsymbol{\psi}_{i-1}\cup\{\psi_m^\star\}
\right).
\end{align}

After selecting $m_i^\star$, the activated segment set is updated as
$\mathcal S_i=\mathcal S_{i-1}\cup\{m_i^\star\}$.
The corresponding achievable sum-rate is then stored. Unlike conventional greedy methods, the proposed algorithm does not terminate once the sum-rate decreases. Instead, all activation levels generated during the greedy search are evaluated, and the best one is finally selected. Since the full-activation case $\mathcal S=\mathcal M$ is included as the last candidate, the proposed Type-I HSS/A is guaranteed to achieve a sum-rate no lower than that of full Type-I SA under the same sequential PA-placement procedure.

The entire procedure is summarized in Algorithm~\ref{alg:typeI}. The computational complexity is dominated by the greedy segment search and the one-dimensional PA placement optimization, which scales as $\mathcal O(M^2KQ)$.

\begin{algorithm}[!t]
\algsetup{linenosize=\tiny}
\scriptsize
\caption{Greedy Type-I HSS/A}
\label{alg:typeI}
\begin{algorithmic}[1]
\REQUIRE $\mathcal M$, $\{P_k\}$, $\sigma^2$, $Q$
\ENSURE $\mathcal S^\star$, $\boldsymbol{\psi}^\star$
\STATE Initialize $\mathcal S_0=\emptyset$, $\boldsymbol{\psi}_0=\emptyset$
\FOR{$i=1$ to $M$}
    \FOR{each $m\in\mathcal M\setminus\mathcal S_{i-1}$}
        \STATE Find $\psi_m^\star$ through one-dimensional search
        \STATE Compute the trial sum-rate after adding segment $m$
    \ENDFOR
    \STATE Select the segment with the largest trial sum-rate
    \STATE Update $\mathcal S_i$ and $\boldsymbol{\psi}_i$
    \STATE Store the achieved sum-rate
\ENDFOR
\STATE Return the activation level with the maximum stored sum-rate
\end{algorithmic}
\end{algorithm}

\subsection{Type-II HSS/A}
For Type-II HSS/A, the phase shifts must be jointly optimized with segment activation and PA placement. For given $\mathcal S$ and $\boldsymbol{\psi}$, the phase-shift design problem can be formulated as follows:
\begin{align}
\max_{\{\theta_m\}}
~
\sum_{k=1}^{K}
P_k
\left|
\sum_{m\in\mathcal S}
{\rm e}^{{\rm j}\theta_m}
g_{k,m}(\psi_m)
\right|^2,
\label{Phase_Problem}
\end{align}
where $\theta_m\in(0,2\pi]$ for all $m\in\mathcal S$.

Define $A_{m,n}
\triangleq
\sum_{k=1}^{K}
P_k
g_{k,m}(\psi_m)
g_{k,n}^*(\psi_n)$ for $m,n\in\mathcal S$ and ${\mathbf v}\triangleq
[{\rm{e}}^{{\rm{j}}\theta_m}]_{m\in\mathcal S}$. Then, \eqref{Phase_Problem} can be rewritten as follows:
\begin{align}
\max_{\mathbf v}
~
\mathbf v^{\mathsf T}
\mathbf A
{\mathbf v}^{*},
\quad
{\rm s.t.}
\quad
|v_m|=1,
~\forall m\in\mathcal S,
\label{Unit_Modulus_Problem}
\end{align}
where $v_m\triangleq{\rm{e}}^{{\rm{j}}\theta_m}$ and $\mathbf A=[A_{m,n}]$. Problem \eqref{Unit_Modulus_Problem} is non-convex due to the unit-modulus constraints. We therefore adopt an element-wise alternating optimization approach. For fixed $\{v_n\}_{n\neq m}$, the optimal update of $v_m$ is given by
\begin{align}
v_m
=
{\rm e}^{-{\rm j}\arg
\left(
\sum_{n\in\mathcal S,n\neq m}
A_{m,n}v_n
\right)}.
\label{Phase_Update}
\end{align}
The above update monotonically improves the objective value in \eqref{Unit_Modulus_Problem} and converges to a stationary point.

\begin{algorithm}[!t]
\algsetup{linenosize=\tiny}
\scriptsize
\caption{Greedy Type-II HSS/A}
\label{alg:typeII}
\begin{algorithmic}[1]
\REQUIRE $\mathcal M$, $\{P_k\}$, $\sigma^2$, $Q$
\ENSURE $\mathcal S^\star$, $\boldsymbol{\psi}^\star$, $\boldsymbol{\theta}^\star$
\STATE Initialize $\mathcal S_0=\emptyset$, $\boldsymbol{\psi}_0=\emptyset$
\FOR{$i=1$ to $M$}
    \FOR{each $m\in\mathcal M\setminus\mathcal S_{i-1}$}
        \STATE Find $\psi_m^\star$ through one-dimensional search
        \STATE Update phase shifts through element-wise alternating optimization
        \STATE Compute the resulting sum-rate
    \ENDFOR
    \STATE Select the segment with the largest sum-rate
    \STATE Update $\mathcal S_i$, $\boldsymbol{\psi}_i$, and $\boldsymbol{\theta}_i$
    \STATE Store the achieved sum-rate
\ENDFOR
\STATE Return the activation level with the maximum stored sum-rate
\end{algorithmic}
\end{algorithm}

The proposed Type-I greedy activation procedure can be naturally extended to Type-II HSS/A. Specifically, for each candidate segment, the PA position is first optimized through one-dimensional search. Then, the phase shifts are updated according to \eqref{Phase_Update}. The resulting sum-rate is evaluated after phase alignment. Similar to Type-I HSS/A, all activation levels generated during the greedy search are stored, and the best one is finally selected. Therefore, the proposed Type-II HSS/A also guarantees a sum-rate no lower than that of full Type-II SA under the same sequential design framework.

The complete procedure is summarized in Algorithm~\ref{alg:typeII}. Compared with Type-I HSS/A, Type-II HSS/A achieves stronger coherent combining gain through analog phase alignment, but it requires additional phase-shifter hardware. The computational complexity mainly arises from the greedy search, PA placement optimization, and iterative phase updates, which scales as $\mathcal O(M^2KQ+I_\theta M^3)$, where $I_\theta$ denotes the number of alternating-optimization iterations for optimizing the phase shifts.

\section{Numerical Results}
This section evaluates the proposed HSS/A schemes. Unless otherwise specified, we set $f_c=28$ GHz, $n_{\mathrm{eff}}=1.4$, $\Delta=\lambda/2$, $d=3$ m, $L=1$ m, $P_k=10$ dBm for all $k\in\mathcal K$, and $\sigma^2=-90$ dBm. The grid-search resolution is set to $Q=10^3$. The users are uniformly distributed within a rectangular region with $D_x=20$ m and $D_y=20$ m, while the waveguide is centered above the service region. All results are averaged over independent user realizations.

\begin{figure}[!t]
\centering
    \subfigure[$K=1$.]
    {
        \includegraphics[height=0.17\textwidth]{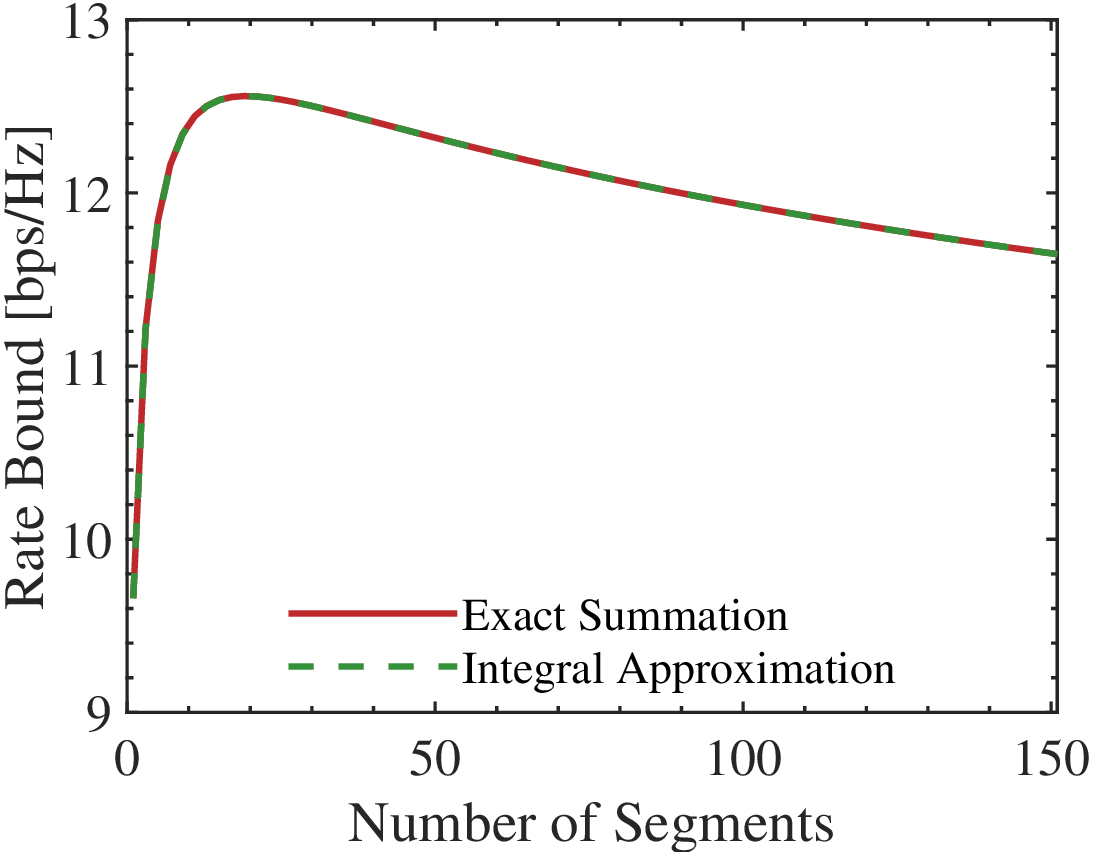}
        \label{Figure: Fig_Rate_Segments}
    }
    \subfigure[$K=4$.]
    {
        \includegraphics[height=0.17\textwidth]{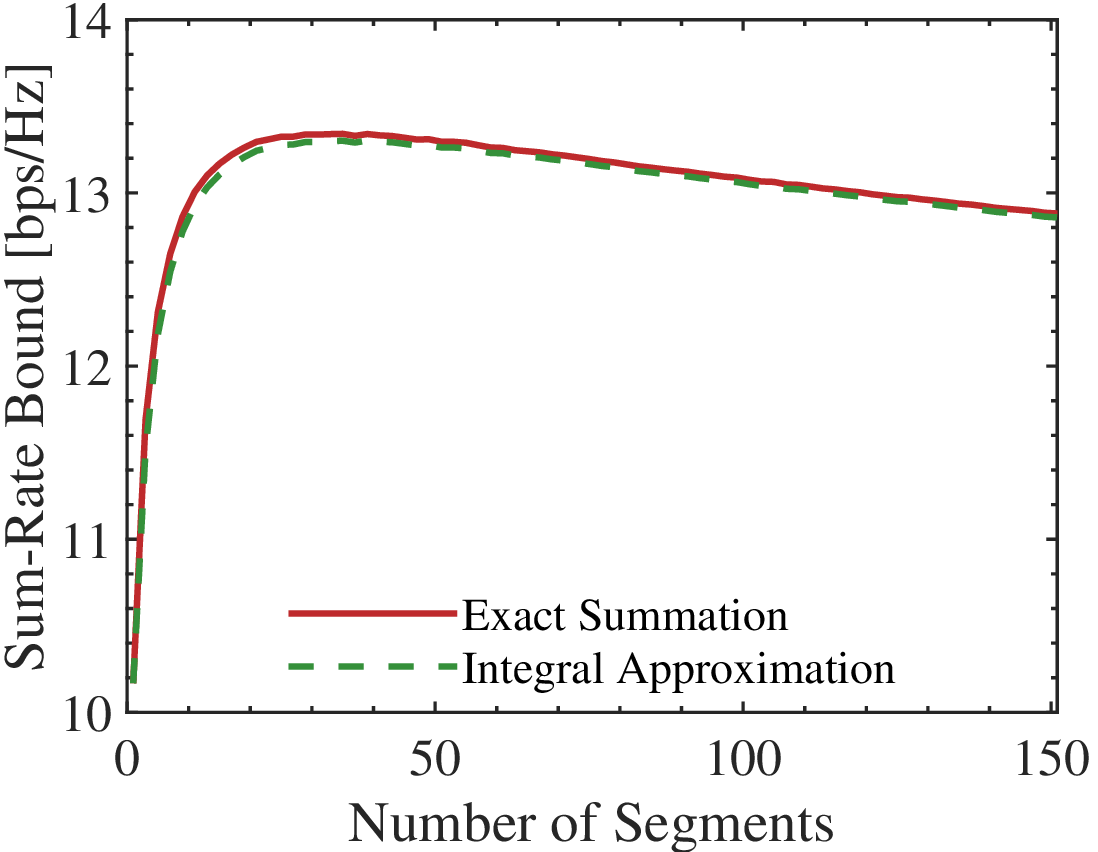}
        \label{Figure: Fig_Rate_Segments1}
    }
\caption{Upper bound of the rate versus the number of segments.}
\label{Figure: Rate_Segments}
\vspace{-10pt}
\end{figure}

Fig.~\ref{Figure: Rate_Segments} validates the derived upper bound and its integral approximation. For both the single-user case and the multiuser case, the integral approximation obtained from \eqref{User_Channel_Gain_Upper_Bound} closely matches the exact summation obtained from \eqref{Channel_Bound_Triangle}. The rate bound first increases and then decreases as the number of segments grows. This observation verifies the analysis in Section~\ref{Section: Sum-Rate Analysis} and confirms that full-segment activation is generally suboptimal. The reason is that nearby segments provide strong coherent combining gain, whereas distant segments contribute only weak signal components while still increasing the accumulated receiver noise.

\begin{figure}[!t]
\centering
    \subfigure[$P_k=-10$ dBm.]
    {
        \includegraphics[height=0.17\textwidth]{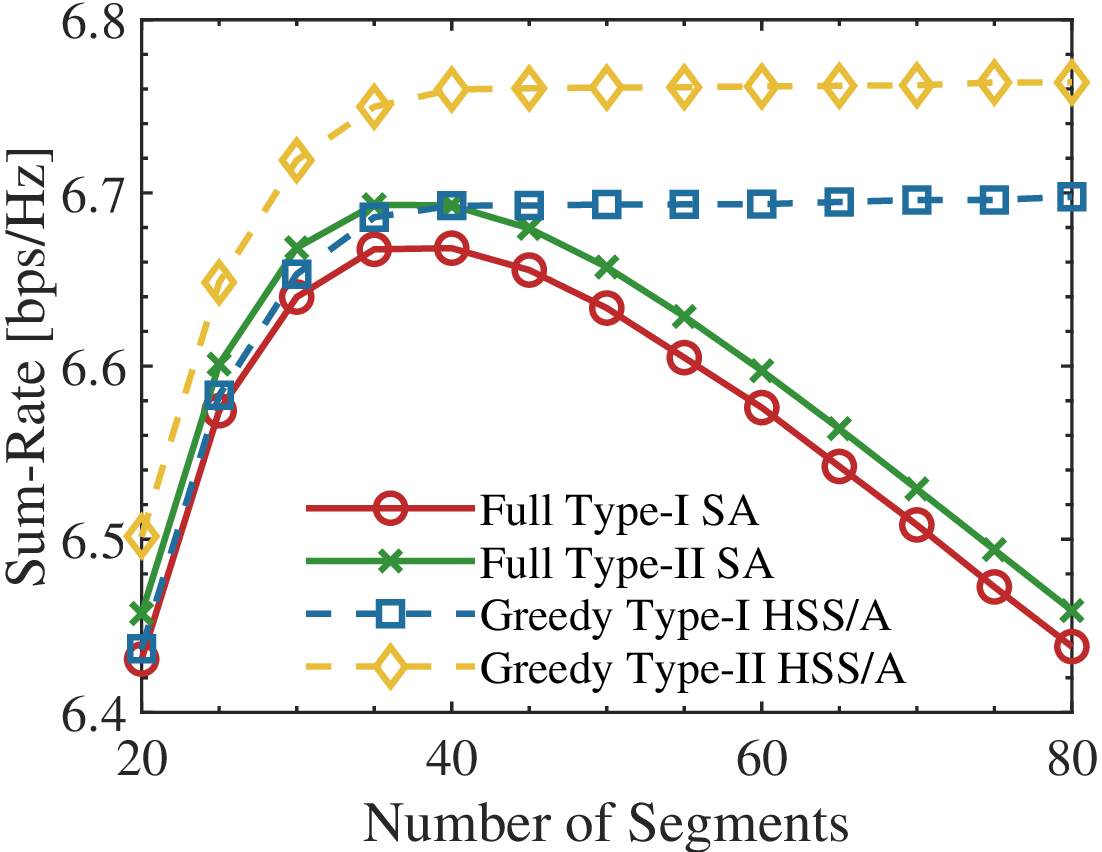}
        \label{Figure: Fig_Segment1}
    }
    \subfigure[$P_k=10$ dBm.]
    {
        \includegraphics[height=0.17\textwidth]{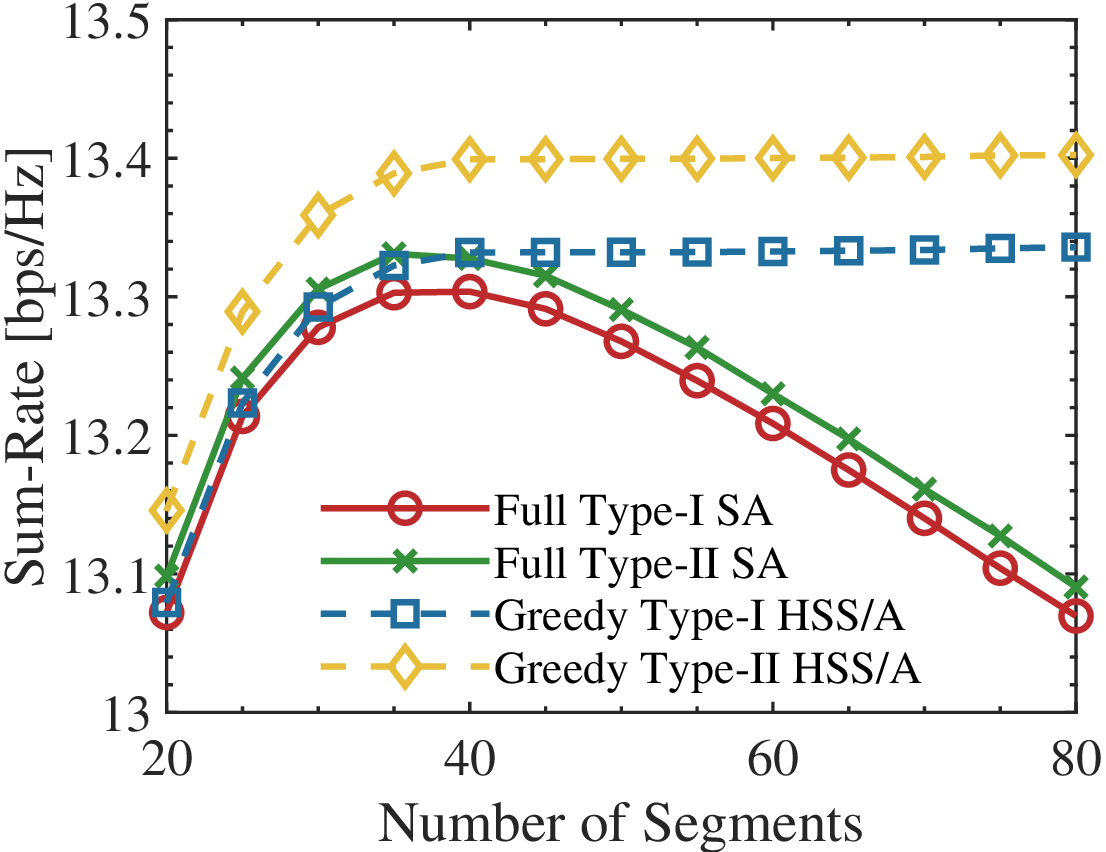}
        \label{Figure: Fig_Segment}
    }
\caption{Sum-rate versus the number of segments. $K=4$.}
\label{Figure: Fig_Segment_Rate}
\vspace{-10pt}
\end{figure}

Fig.~\ref{Figure: Fig_Segment_Rate} compares full SA and the proposed HSS/A schemes. For full SA, the sum-rate is obtained through joint PA-placement and phase-shift optimization, where the PA locations are refined by element-wise alternating optimization with one-dimensional search, while the phase shifts are updated according to \eqref{Phase_Update}. The sum-rate of full SA first increases and then decreases with $M$, which agrees with the theoretical trend shown in Fig.~\ref{Figure: Rate_Segments}. By contrast, HSS/A stores all activation levels generated during the greedy search and finally selects the best one. Therefore, its sum-rate does not decrease when additional available segments are deployed. The proposed greedy HSS/A schemes outperform their full-SA counterparts under both transmit-power settings. This gain is achieved by avoiding harmful segments whose marginal signal contribution is insufficient to compensate for the additional receiver noise. Type-II HSS/A further improves the sum-rate through analog phase alignment across the activated segments. However, the improvement remains moderate because one phase shift is shared by all users within each segment and cannot simultaneously align the received signals of all users.

\begin{figure}[!t]
  \centering
  \includegraphics[height=0.18\textwidth]{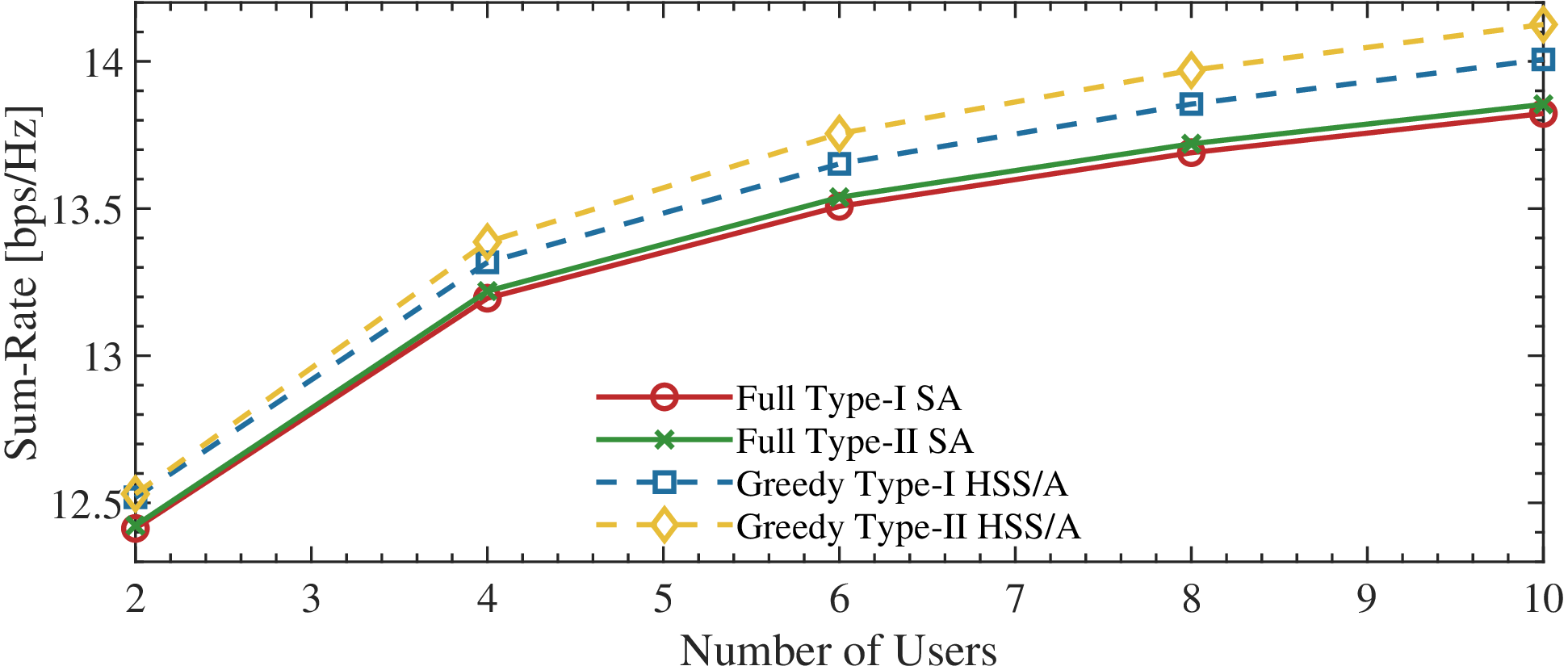}
  \caption{Sum-rate versus the number of users. $M=60$.}
  \label{Fig_User}
\vspace{-10pt}
\end{figure}

Fig.~\ref{Fig_User} shows the impact of the number of users when $M=60$. The achievable sum-rate increases with $K$ because more users contribute to the uplink NOMA transmission. The proposed HSS/A schemes consistently outperform full SA, which confirms that segment activation remains beneficial in multiuser scenarios. Type-II schemes achieve higher rates than Type-I schemes due to the additional phase-control capability. Nevertheless, the performance gap is limited because the phase shifts are optimized for the overall multiuser objective rather than for each user individually.

\section{Conclusion}
This article proved that activating more segments in SWAN does not always improve the achievable sum-rate due to accumulated receiver noise. Motivated by this observation, we developed low-complexity greedy HSS/A schemes for joint segment activation and PA placement. The obtained results provide useful insights for the design of large-scale SWAN systems.

\bibliographystyle{IEEEtran}
\bibliography{mybib}
\end{document}